\documentclass[onecolumn,american,english]{sig-alternate}
\usepackage[T1]{fontenc}
\usepackage[latin9]{inputenc}
\usepackage{babel}
\usepackage{array}
\usepackage{multirow}
\usepackage{amsmath}
\usepackage{amssymb}
\usepackage{graphicx}
\usepackage[unicode=true,
 bookmarks=true,bookmarksnumbered=false,bookmarksopen=false,
 breaklinks=false,pdfborder={0 0 0},pdfborderstyle={},backref=page,colorlinks=false]
 {hyperref}
\hypersetup{pdftitle={Work in Progress: Learning Term Weights for Ad-hoc Retrieval},
 pdfauthor={B. Piwowarski}}

\makeatletter

\providecommand{\tabularnewline}{\\}

\usepackage{tikz}
\usetikzlibrary{matrix,calc,positioning,shapes}

\makeatother

\begin{document}

\title{Learning Term Weights for Ad-hoc Retrieval}

\author{B. Piwowarski (UPMC/CNRS, Paris, France)\\
{\footnotesize{}benjamin@bpiwowar.net}}
\maketitle
\begin{abstract}
Most Information Retrieval models compute the relevance score of a
document for a given query by summing term weights specific to a document
or a query. Heuristic approaches, like TF-IDF, or probabilistic models,
like BM25, are used to specify how a term weight is computed. In this
paper, we propose to leverage learning-to-rank principles to learn
how to compute a term weight for a given document based on the term
occurrence pattern.
\end{abstract}

\section{Introduction}

\begin{sloppypar}Ad-hoc Information Retrieval aims at ranking documents
according to their relevance to a query. Many different models exist,
such as BM25 \cite{Robertson2009The-Probabilistic-Relevance} and
language models \cite{Zhai2008LMOverview}. The core of most IR model
is the term weighting formula, that assigns a weight to each term
of each document according to its importance \textendash{} how likely
the document is relevant if such a term appears in a query? Such term
weighting functions are crucial, even for state of the art learning-to-rank
approaches \cite{Liu2011LETOR} \textendash{} the fact that such approaches
systematically use one or more IR models as term weight-based features
outlines their importance.

Term weighting functions have been, from the very beginning of research
on IR models, a focus of many works. Attempts to improve the weighting
scheme include developing different models of the IR process \cite{Zhai2008LMOverview,Amati:2002tm},
estimating more reliably some of its components \cite{LeZhao2010TermNecessity},
or, and this is the focus of this paper, learning the term weighting
function \cite{Taylor2006Optimisation}. As the latter leverages the
same source of information as learning-to-rank models, learning the
term weight function has a great potential, provided enough training
data, and expressiveness power for the function.

The approach we propose is inspired by recent work in representation
learning \cite{Bengio2013RepresentationOverview} whose main idea
is to process directly raw data rather than computing features. In
Information Retrieval, this corresponds to designing a model that
would take as input words, outputting a score for a given document.
This approach has been followed by some recent works like \cite{Palangi2015LSTMIR},
using recurrent neural networks, and \cite{Huang:2013wk}, using convolutional
neural networks. Those works aim at embedding documents in a low-dimensional
space, and are thus comparable to latent space approaches like LSA
\cite{Deerwester:1990gu}, sharing their properties \cite{Hoenkamp:2011gf}
of increasing recall at the expense of precision. This has been confirmed
in tasks like question-answering, where a query referring to a named
entity should be answered by documents containing it. In that case,
approaches based on dimensionality reduction do not perform well,
even if trained in a supervised way \cite{Severyn:2015gm}. Another
drawback of such approaches is that they need great quantities of
data to set the different parameters of these models, since rare terms
will tend to have their representation badly estimated. As rare terms
might be good indicators of document relevance, we believe this is
a problem.

The main proposal of this paper is that a representation of a term
be the positions it occurs in in the documents of the collection and
in the document for which we want to compute the term weight. We can
then compare terms based on their \emph{patterns of occurrences }to
determine what their term weight should be. The advantages of doing
so, compared to actual neural network based approaches, is that (1)
less training data is needed since two terms might share a common
occurrence pattern while being semantically not related; (2) inverted
indices can still be used, allowing fast retrieval. Finally, compared
to standard IR weighting schemes, no prior hypothesis on the functional
form of the term weighting function is made.

Using term occurrence pattern would help to distinguish the cases
of \textendash{} the list is not exhaustive:
\begin{enumerate}
\item Terms occurring regularly, whatever the document, would likely be
unimportant words.
\item Terms occurring most of the time at the same position (e.g. ``Introduction''
for scientific papers) could be important when occurring at other
places.
\item Terms occurring throughout a document would be more important than
those occurring only in one part.
\end{enumerate}
In this paper, we present preliminary work that we have undertaken
in this direction, using a representation based on a clustering of
the term occurrence patterns. In the following, we first present related
work (Section 2), before introducing our approach (Sections 3 and
4), giving some preliminary experimental results (Section 5), and
concluding (Section 6).

\section{Related Work}

Related work can be divided in two parts: (1) some works have tried
to learn various parts of the term weighting model \textendash{} ranging
from hyperparameters to full term weighting functions; (2) more recently,
some approaches have tried to compute the relevance score of a document
for a given query, using neural networks. We describe those two lines
of research in turn.

Taylor et al. \cite{Taylor2006Optimisation} proposed to learn the
BM25 parameters ($k_{1}$ and $b$) using a pairwise RankNet loss
function, and have shown that this led to the best results achievable
given appropriate training. Schuth et al. \cite{Schuth2014OptimizingBM25}
extended this approach by leveraging user clicks rather than relevance
assessments. 

Rather than relying on a pre-established term weighting function whose
hyperparameters are learned, it can be interesting to learn the term
weighting function directly. This has been explored by using genetic
algorithms to evolve term weighting functions (represented as trees)
\cite{Cummins2006evolving}, using terminals like term frequency and
document frequency, and non-terminals (functions) like sum, product
and logarithm. Closer to our work, Svore et al. \cite{Svore2009LTW}
proposed to use a multi-layer neural network to learn directly the
term weight given features like term frequency and document frequency.
In this work, we go further, and instead of using pre-defined features,
we propose to estimate the term weight directly from the occurrences
of a given term in the document and its context (the document collection). 

This approach goes in the same direction as deep neural architectures
that have had a great success in the field of image processing \cite{Bengio2013RepresentationOverview},
one of the working hypothesis is to train a neural network to predict
a value given the raw representation of an object. In the field of
natural language processing and information access, this usually means
that terms are used as the input. Most models rely on low-dimensional
representation of words, such as those obtained by word2vec \cite{Mikolov2013Word2vec},
where each word is associated to one vector in a low-dimensional latent
space.

Huang et al. \cite{Huang:2013wk} proposed to use a convolutional
neural network. Instead of starting from word embeddings, they did
use letter tri-grams (so as to have a small size vocabulary), i.e.
each document is represented by the count of the tri-grams occuring
in it. The output is a fixed size representation vector that is used
to compute the relevance to a query. Shen et al. \cite{Shen:2014id}
extended their work by first computing the representation of word
tri-grams, before using a \emph{pooling layer }(maximum over each
dimension of the representation space) to represent the full document.
Finally, \cite{Palangi2015LSTMIR} used a recurrent neural network
(RNN), the representation of a document or a query being the state
of the RNN at the end of the processed sequence. Compared to our work,
these approach need a great quantity of training data, and we believe
they are not suited for many IR tasks dealing with precise named entities.
In the context of question-answering, \cite{Severyn:2015gm} proposed
to learn whether a sentence is an answer to a given query using a
convolutional neural network, but had to introduce a set of query-document
features to improve their results, such as the word overlap count.

In parallel, Zheng and Callan \cite{Zheng2015LearningToReweight}
proposed a \emph{somehow} term-independent representation of query
terms to define the \emph{query} weight of each term. The central
idea of their work is to represent each term of the query as the difference
between the term vector and the mean of the vectors representing the
terms of the query thus capturing the semantic difference between
the term and the query. Our research is direction is orthogonal, since
we are interested by the document weight and not the query one, but
the idea of finding a term-independent representation inspired our
present work. 

\section{Problem formulation}

We start by exposing briefly the overall learning-to-rank optimization
scheme. The relevance score of a document for a given query is given
as a weighted sum, over terms present in the query, of their importance
$w\left(t,d\right)$, that is

\begin{equation}
s_{\theta}\left(q,d\right)=\sum_{t\in q}f_{\theta}\left(t,q\right)w_{\theta}\left(t,d\right)\label{eq:mainrsv}
\end{equation}
where $f_{\theta}(t,q)$ denotes the importance of the term(s) $t$
in the query $q$ (we suppose in this paper that it is a constant
equal to 1) and $w_{\theta}\left(t,d\right)$ is the computed term
weight. Both depend on the model parameters $\theta$ . 

To learn the parameters $\theta$, many different optimization functions
could be used. We choose the RankNet pairwise criterion \cite{Burges:2005ts}
because it is simple and was shown to perform well on a variety of
test collections. It supposes that the probability that a document
$a$ is ranked before the document $b$ given a query $q$ is given
by
\[
\sigma\left(s_{\theta}(q,a)-s_{\theta}(q,b)\right)
\]
where $\sigma\left(x\right)$ is the sigmoid function $1/\left(1+\exp(-x)\right)$.
The cross-entropy cost function is then used to optimize the parameters
$\theta$ of the model. In our case, this gives

\[
\mathbb{E}\left(\sigma\left(\sum_{t\in q}w_{\theta}(t,a)-w_{\theta}\left(t,b\right)\right)\right)
\]
where the expectation is over all triplets $(q,a,b)$ such that $a$
is more relevant than $b$ \textendash{} for binary relevance like
in our experiments, this corresponds to the cases where $a$ is relevant
and $b$ is not. We further suppose that $w_{\theta}(t,d)=0$ if the
term $t$ does not appear in the document $d$ \textendash{} this
is a common hypothesis made by all term weighting schemes.

\selectlanguage{american}%
\begin{center}
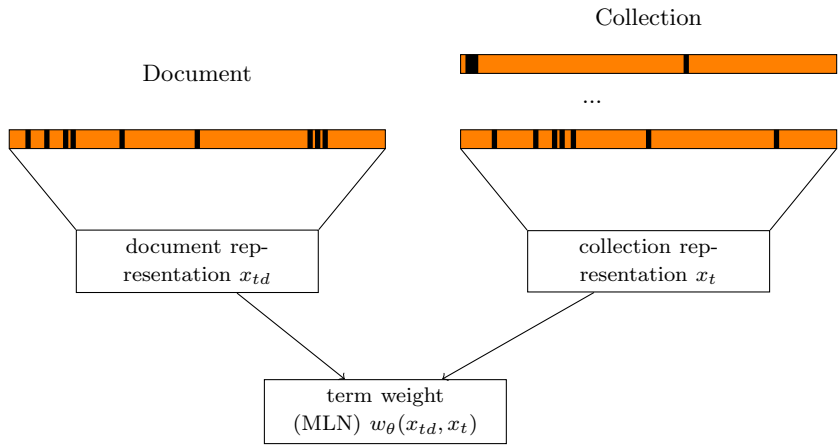
\begin{figure*}
\begin{centering}
\begin{tikzpicture}[scale=0.5]


\node[align=center] at (17, 3.5) {Collection};
\draw[fill=orange] (12, 2) rectangle ++(10, 0.5);
\foreach \x in {12.2, 12.3, 12.4, 18} {
    \draw[line width=2] (\x,2) -- (\x,2.5);
}
\draw (15.5, 1.25) node {...};
\draw[fill=orange] (12, 0) rectangle ++(10, 0.5);
\foreach \x in {12.9, 14, 14.5, 14.7, 15, 17, 20.4} {
    \draw[line width=2] (\x,0) -- (\x,.5);
}

\node[rectangle,draw,text centered,text width=3cm] (collrep) at (17, -3) {\small collection representation $x_{t}$};
\draw (12, 0) -- (collrep.north west);
\draw (22, 0) -- (collrep.north east);


\draw[fill=orange] (0, 0) rectangle ++(10, 0.5);
\node[align=center] at (5, 2) {Document};
\foreach \x in {0.5, 1, 1.5, 1.7, 3, 5, 8, 8.2, 8.4} {
    \draw[line width=2] (\x,0) -- (\x,0.5);
}


\node[rectangle,draw,text centered,text width=3cm] (docrep) at (5, -3) {\small document representation $x_{td}$};

\draw (0, 0) -- (docrep.north west);
\draw (10, 0) -- (docrep.north east);


\node[rectangle,draw,text centered,text width=3cm] (weight) at (10, -7) {\small term weight (MLN) $w_\theta(x_{td}, x_t)$};

\draw[->] (docrep) -- (weight);
\draw[->] (collrep) -- (weight);

\end{tikzpicture}
\par\end{centering}
\caption{\foreignlanguage{english}{\label{fig:general-model}General model for computing the weight of
term $t$ in document $d$. The term occurrences are marked with a
black bar, and are used to compute for a single term (left) the representation
of a document (right) the representation of the collection. Both representations
can then be used to compute the final term weight.}}
\end{figure*}
\par\end{center}

\selectlanguage{english}%
We now turn to the problem of computing $w_{\theta}\left(t,a\right)$.
The general model is presented in Figure\,\ref{fig:general-model},
where the weight of a term is computed by considering two sources
of information, the document (left) and the collection (right). We
would like to be able to compute a term-document representation $x_{td}\in\mathbb{R}^{n}$
that \emph{captures} the pattern of occurrence of term\,$t$ in document\,$d$,
and of a term-collection representation $x_{tc}\in\mathbb{R}^{p}$
that \emph{captures }the pattern of occurrence of term\,$t$ in collection\,$c$.
Given these two pieces of information, a term weight can be computed
through the function $w_{\theta}$. 

\section{Proposed model}

While we could learn directly the term weight given the term occurrence
pattern in the document and the collection, we choose to decompose
the problem in two parts: (1) computing a faithful representation
of the document/collection term occurrence pattern and (2) computing
the term weight. End-to-end learning is a longer term objective, but
we first need to find the most promising options. While other choices
are possible and will be explored, we describe next a first and simple
instance of this model, using K-Means for (1) and neural networks
for (2). 

\subsection{Document and collection representation with K-means}

\label{subsec:k-means}

In this preliminary work, we experimented with a simple method where
we used $k$-means to compute the term document and collection representations.
We followed a standard approach for representing object with $k$-means
\cite{Coates2012KMeansRep}: we first clustered patterns of occurrences
of all the terms into $k$ clusters; then, each document is represented
by a vector in $\mathbb{R}^{k}$ where each component is the distance
of the corresponding centroid to the document representation. We now
describe more in details the methodology.

The initial document representation is a probability distribution
over the positions in a document, i.e. the probability of a term $t$
occurring at a given position $p$ in document $d$ is given by:

\texttt{
\[
P\left(p|d\right)=\frac{1}{\#d}\delta_{pd}
\]
}where $\#d$ is the number of term occurrences in documents (i.e.
its length), $\delta_{pd}$ is the Dirac function that is 1 when the
term appears at position $p$. The position $p$ can be \emph{normalized}
or not: if normalized, $p$ ranges from 0 to 1, otherwise from 0 to
the length of the document. The interest of normalizing is that documents
of different lengths can be directly compared, and more meaningful
clusters be found.

To compute the distance between two document distributions, we use
the quadratic Wasserstein metric $W_{2}$ \cite{Rabin2011Wasserstein}
which amounts at computing the minimum ``move of probability mass''
(squared distance $\times$ probability mass) so that one (continuous)
distribution can be transformed in another \textendash{} it is related
to the Earth Mover Distance (EMD) \textemdash{} using the square of
the distance rather than the distance itself, and has interesting
computational properties for k-means. This distance seems also more
adapted than a $L_{2}$ distance (in the vector space defined by quantized
positions) since it takes into account the proximity of one position
to another.

Unfortunately, for computational reasons, we had to represent the
distribution by vectors of fixed size (we used a dimension $D=$100
in the experiments), i.e. a document is represented as a vector $p_{d}\in\mathbb{R}^{D}$.
The distribution probability can then be defined from the vector $p_{d}$
as:
\[
\tilde{P}\left(p|d\right)=\frac{1}{d}\#\left\{ j|p_{dj}=p\right\} 
\]
which is an approximation of the original probability \textendash{}
the $p_{dj}$ were computed so as to minimized the $W_{2}$ distance
between $\tilde{P}$ and $P$. If the $p_{dj}$ are increasing (i.e.
$p_{dj}\ge p_{d_{j^{\prime}}}$ if $j\ge j^{\prime}$), the $L_{2}$-norm
can be used directly to compute the $W_{2}$ distance between two
distributions while the mean can be used to compute the centroid \cite{Rabin2011Wasserstein}.

The representation $x_{td}$ of a term for a document is then given
by its distance to the corresponding centroid, that is the $i$th
component of the vector representing the term $t$ in document $d$
is given by:

\[
x_{tdi}=W_{2}\left(p_{d},c_{i}\right)
\]
where $c_{i}$ is the $i$th cluster.

For the term-collection representation $x_{t}$, we chose to either
compute the mean of the document vectors for the corresponding terms,
or to compute the sum. The interest of the latter is that the sum
captures somehow the number of occurrences of a term in the document
collection. Formally,
\[
x_{t}^{(sum)}=\sum_{d/t\in d}x_{td}
\]
 and

\[
x_{t}^{(mean)}=\frac{1}{\#\left\{ d/t\in d\right\} }\sum_{d/t\in d}x_{td}
\]

\subsection{Term weighting}

\label{subsec:term-weighting}

Knowing $x_{t}$ and $x_{td},$ we then used a multi-layer network
to compute the score of a term given its document and collection representation.
Each layer was composed by a linear transformation followed by an
activation function \textendash{} we choose the ReLU activation \cite{Nair2010Relu},
which was shown in many cases to facilitate learning. Each layer corresponds
to the function
\[
y=\max\left(0,Ax+b\right)
\]
where $x$ is the input, $y$ the output, $A$/$b$ the parameters,
and the maximum is component-wise. The first layer had $2k$ inputs,
corresponding to the size of the vectors $x_{t}$ and $x_{td}$. The
last layer has one output, which corresponds to the term weight. In
this work, we only experimented with one hidden layer (of size 50).

\section{Preliminary Experiments}

\emph{Test collections.} We used the TREC 1 to 8 collections. We split
the dataset in two parts (TREC 1-4, TREC 5-8), and use one part to
train the models (TREC 1-4, resp. TREC 5-8), and the other (TREC 5-8,
resp. TREC 1-4) to evaluate its performance using mean average precision
(MAP). Experiments were conducted using the \emph{title }field of
the TREC topics (except for TREC-4 where only long versions are available,
and were thus used).

\emph{Models.} In this preliminary set of experiments, we compared
three models \textendash{} it would be useful to compare the results
to neural networks approaches trained on the same dataset to get a
full picture of the method potential, but in this preliminary work
we were more interested in bringing our model up to the standard IR
models:
\begin{enumerate}
\item BM25 with hyperparameters set to their default values, i.e $k_{1}=1.2$
and $b=0.75$ 
\item BM25 whose hyperparameters $k_{1}$ and $b$ were learnt, following
the work described in \cite{Taylor2006Optimisation}. The weighting
function is
\begin{multline*}
w(t,d)=\frac{tf(t,d)\cdot(k_{1}+1)}{tf(t,d)+k_{1}\left(1-b+b\cdot\frac{l_{d}}{\bar{l}}\right)}\\
\times\log\frac{N-df(t)+0.5}{df(t)+0.5}
\end{multline*}

where $k_{1}$ that controls the bustiness of words (how likely a
term occur again) and $b$ that controls the verbosity of documents
(are the documents mono or multi-topical?). We left out the parameter
$k_{3}$ of our study since it is not useful when query terms occur
only once (in queries).
\item K-Means + MLN as described in Sections \ref{subsec:k-means} (K-Means)
and \ref{subsec:term-weighting}. We experimented with the following
parameters:
\begin{enumerate}
\item Dimension of the document and collection representations was varied,
taking the values $K=$10, 50, 100 and 200.
\item The position was normalized or not (i.e. divided or not by the length
of the document).
\item The MLP hidden layer was composed of 50 units
\end{enumerate}
\end{enumerate}
Both models were trained with the ADAM optimizer that takes into account
second order information \cite{Kingma:2014us}, with $\epsilon$ set
to respectively 1e-4 (BM25) and 1e-8 (our model), and other hyperparameters
set to their default values. They were found to decrease the training
cost in our preliminary experiments. We used 1000 iterations for learning,
where during each iterations we computed the loss with respect to
a sample of 50\,000 triples (query, document $a$ and $b$).

\begin{table*}
\begin{centering}
\begin{tabular}{|c|c||c|c|c|c||c|}
\cline{1-6} 
\multirow{2}{*}{BM25} & \multirow{2}{*}{BM25+} & \multicolumn{2}{c|}{Not Normalized} & \multicolumn{2}{c|}{Normalized} & \multicolumn{1}{c}{}\tabularnewline
\cline{3-7} 
 &  & Mean & Sum & Mean & Sum & $k$\tabularnewline
\hline 
\hline 
\multirow{4}{*}{0.18} & \multirow{4}{*}{0.19} & 0.03 & 0.03 & 0.10 & 0.11 & 10\tabularnewline
\cline{3-7} 
 &  & 0.06 & 0.07 & 0.10 & 0.10 & 50\tabularnewline
\cline{3-7} 
 &  & 0.07 & 0.09 & 0.11 & 0.09 & 100\tabularnewline
\cline{3-7} 
 &  & 0.10 & 0.09 & 0.11 & 0.09 & 200\tabularnewline
\hline 
\end{tabular}
\par\end{centering}
\caption{\label{tab:results}Results for all models \textendash{} averaged
over all TRECs}
\end{table*}

\begin{figure}
\begin{centering}
\includegraphics[width=1\columnwidth]{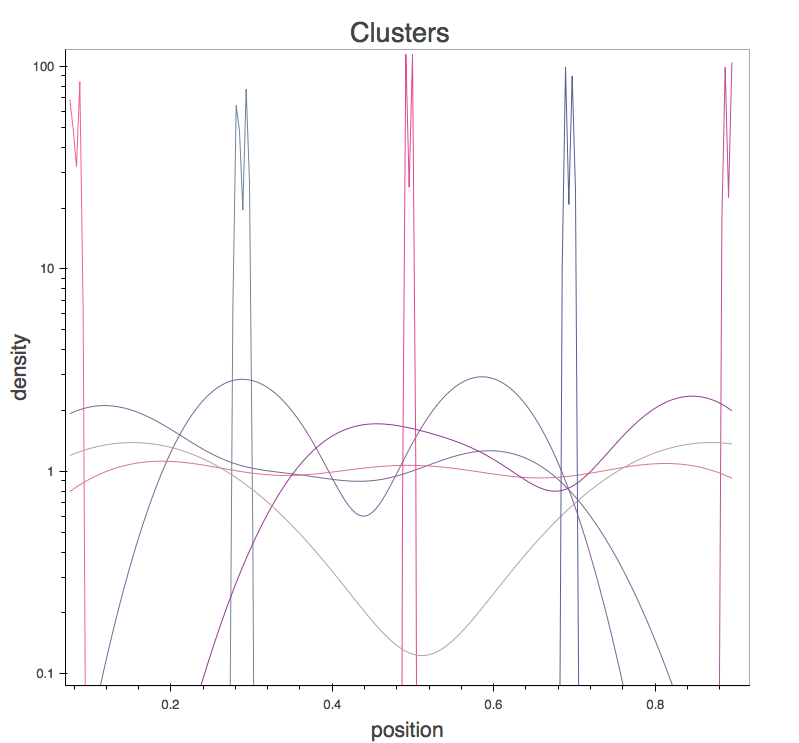}
\par\end{centering}
\caption{\label{fig:clusters-10}Discovered clusters ($k=10$, normalized length)}
\end{figure}

We report the results in Table \ref{tab:results}. As the behavior
of the different systems was stable over these collections, we average
the MAP over TREC-1 and TREC-8. Experimental results confirm the fact
that learning the BM25 parameters using a learning-to-rank cost function
is effective \cite{Taylor2006Optimisation}. When comparing the performance
of BM25 to our proposed model, the results were disappointing, but
we can formulate the following observations:
\begin{enumerate}
\item We observed that in some cases the learning process did not converge,
so actual results might be higher than presented;
\item Normalized length seem to provide better results than non-normalized
when the dimension is low (10-100) and then the performance is matched
when $k=200$. This is consistent with the fact that when positions
are not normalized, more clusters are necessary to distinguish different
patterns of occurrence in documents of varying lengths. Experiments
will be run with higher dimensions to explore whether this trend holds;
\item Increasing the representation dimension seems to improve results for
all models but the normalized/sum one \textendash{} we have no clear
explanation for this observation;
\item There is no clear effect of using a sum or a mean for the document
collection representation, but looking at the learned parameters will
help in determining the usefulness of each part of the representation.
\end{enumerate}
We also looked at the clusters. In the case of not normalized positions,
the discovered clusters were mostly positions within the document
(i.e. the distribution corresponds to a single Dirac function); this
shows that there was not pattern discovered with such an approach.
Normalized positions led to better clusters as shown in Figure \ref{fig:clusters-10}.
In this case, we have 5 clusters corresponding to specific positions
in a document which are evenly spaced. We have then 5 different clusters,
corresponding for example to terms occurring throughout the document,
or at the beginning and the end. Even in this case, the information
captured might not reflect the true diversity of word occurrences
\textendash{} a better way to compute document and collection term
representations could be necessary.

\section{Conclusion and future work}

In this paper, we proposed to learn the term weight directly from
the occurrences of terms in documents. This has the potential to capture
patterns of occurrences that are linked with the importance of a term
in a document and a collection. Once learned \textendash{} and provided
the model is improved, indexing a full collection could be performed
with a low penalty compared to other weighting schemes like e.g. BM25.

Preliminary results have shown that there is still an important gap
between standard IR models and this type of approach. We believe that
this is due to several reasons, the first one being the poor representation
as computed by k-means with the Wasserstein distance. Besides increasing
the model expressiveness (dimension, size and number of hidden layers
in the MLP), we believe that this is due to the fact that (1) we did
use rough approximations (for position distribution), and (2) the
Wasserstein distance is not really adapted to the IR setting. For
instance, it does not properly model the \emph{burstiness} of a term
\cite{Church2008Poisson} that seems to be an important property of
word occurrences. We are working on a recurrent neural network model
that would maximize the probability of observing a series of positions
in a document: 
\[
P\left(p_{1}|x_{td}\right)\ldots P\left(p_{n}|p_{1},\ldots,p_{n-1},x_{td}\right)P\left(x_{td}|x_{d}\right)
\]
where the probability $P\left(x_{td}|x_{d}\right)$ would model the
distribution probability of the positions of a term/document representation
knowing the term/collection representation. Both vectors $x_{t}$
and $x_{td}$ would be learned, and the various models could be compared
on their likelihood before using one for term weight prediction.

To further strengthen the model, we could furthermore try to integrate
constraints formulated in \cite{Clinchant:2010ja} to regularize or
constrain the functional form of the computation of a term weight.
Semantic proximity between terms could be used to increase the accuracy
of the term weighting function. This could be achieved by encoding
not only the position of a term, but the positions of related terms.

Finally, we believe that this approach could be extended in several
interesting ways: (1) by computing term weights for frequent bi or
tri-grams, in order to capture concepts like ``information retrieval'';
(2) by computing the full RSV of a document given the pattern of occurrence
of the different query terms. In the latter case, we could capture
the fact that some query terms occur close to each other \textendash{}
extending approaches like e.g. positional language models \cite{Lv:2009:PLM:1571941.1571994}.\end{sloppypar}

\bibliographystyle{abbrv}
\bibliography{ltw-sigir2016}

\begin{thebibliography}{10}

\bibitem{Amati:2002tm}
G.~Amati and K.~van Rijsbergen.
\newblock {Probabilistic models of information retrieval based on measuring the
  divergence from randomness}.
\newblock 2002.

\bibitem{Bengio2013RepresentationOverview}
Y.~Bengio.
\newblock {Deep Learning of Representations: Looking Forward}.
\newblock May 2013.

\bibitem{Burges:2005ts}
C.~Burges, T.~Shaked, E.~Renshaw, A.~Lazier, M.~Deeds, N.~Hamilton, and
  G.~Hullender.
\newblock {Learning to rank using gradient descent}.
\newblock pages 89--96, 2005.

\bibitem{Church2008Poisson}
K.~W. Church and W.~A. Gale.
\newblock {Poisson mixtures}.
\newblock {\em Natural Language Engineering}, 1(02):163--190, September 2008.

\bibitem{Clinchant:2010ja}
S.~Clinchant and E.~Gaussier.
\newblock {Information-based models for ad hoc IR}.
\newblock In {\em The International ACM SIGIR Conference}, page 234, New York,
  New York, USA, 2010. ACM Press.

\bibitem{Coates2012KMeansRep}
A.~Coates and A.~Y. Ng.
\newblock {Learning Feature Representations with K-Means}.
\newblock In {\em Neural Networks: Tricks of the Trade}, pages 561--580.
  Springer Berlin Heidelberg, Berlin, Heidelberg, 2012.

\bibitem{Cummins2006evolving}
R.~Cummins and C.~O'Riordan.
\newblock {Evolving local and global weighting schemes in information
  retrieval}.
\newblock {\em Information Retrieval Journal}, 9(3):311--330, 2006.

\bibitem{Deerwester:1990gu}
S.~Deerwester, S.~T. Dumais, G.~W. Furnas, T.~K. Landauer, and R.~Harshman.
\newblock {Indexing by latent semantic analysis}.
\newblock {\em Journal of the American Society for Information Science},
  41(6):391--407, September 1990.

\bibitem{Hoenkamp:2011gf}
E.~Hoenkamp.
\newblock {Trading Spaces: On the Lore and Limitations of Latent Semantic
  Analysis}.
\newblock In {\em Advances in Information Retrieval Theory}, pages 40--51.
  Springer Berlin Heidelberg, Berlin, Heidelberg, 2011.

\bibitem{Huang:2013wk}
P.~S. Huang, X.~He, J.~Gao, L.~Deng, A.~Acero, and L.~Heck.
\newblock {Learning deep structured semantic models for web search using
  clickthrough data}.
\newblock In {\em ACM conference on Conference on Information and Knowledge
  Management}, 2013.

\bibitem{Kingma:2014us}
D.~Kingma and J.~Ba.
\newblock {Adam: A Method for Stochastic Optimization}.
\newblock {\em arXiv.org}, December 2014.

\bibitem{Liu2011LETOR}
T.-Y. Liu.
\newblock {\em {Learning to Rank for Information Retrieval}}.
\newblock Springer Berlin Heidelberg, Berlin, Heidelberg, 2011.

\bibitem{Lv:2009:PLM:1571941.1571994}
Y.~Lv, Y.~Lv, and C.~X. Zhai.
\newblock {Positional language models for information retrieval}.
\newblock In {\em The International ACM SIGIR Conference}, page 299, New York,
  New York, USA, July 2009. ~ACM Request Permissions.

\bibitem{Mikolov2013Word2vec}
T.~Mikolov, I.~Sutskever, K.~Chen, G.~S. Corrado, and J.~Dean.
\newblock {Distributed Representations of Words and Phrases and their
  Compositionality.}
\newblock {\em NIPS'14}, cs.CL:3111--3119, 2013.

\bibitem{Nair2010Relu}
V.~Nair and I.~S.~G. Hinton.
\newblock {Rectified linear units improve restricted boltzmann machines}.
\newblock In {\em International Conference on Machine Learning}, 2010.

\bibitem{Palangi2015LSTMIR}
H.~Palangi, H.~Palangi, L.~Deng, Y.~Shen, J.~Gao, X.~He, J.~Chen, X.~Song, and
  R.~Ward.
\newblock {Deep Sentence Embedding Using the Long Short Term Memory Network:
  Analysis and Application to Information Retrieval}.
\newblock {\em arXiv.org}, February 2015.

\bibitem{Rabin2011Wasserstein}
J.~Rabin, G.~Peyr{\'e}, J.~Delon, and M.~Bernot.
\newblock {Wasserstein Barycenter and Its Application to Texture Mixing.}
\newblock {\em SSVM}, 6667(Chapter 37):435--446, 2011.

\bibitem{Robertson2009The-Probabilistic-Relevance}
S.~E. Robertson and H.~Zaragoza.
\newblock {\em {The Probabilistic Relevance Framework: BM25 and Beyond}},
  volume~3 of {\em Foundations and Trends in Information Retrieval}.
\newblock 2009.

\bibitem{Schuth2014OptimizingBM25}
A.~Schuth, F.~Sietsma, S.~Whiteson, and M.~de~Rijke.
\newblock {Optimizing Base Rankers Using Clicks - A Case Study Using BM25.}
\newblock {\em ECIR}, 8416(Chapter 7):75--87, 2014.

\bibitem{Severyn:2015gm}
A.~Severyn and A.~Moschitti.
\newblock {Learning to Rank Short Text Pairs with Convolutional Deep Neural
  Networks}.
\newblock In {\em The International ACM SIGIR Conference}, pages 373--382, New
  York, New York, USA, 2015. ACM Press.

\bibitem{Shen:2014id}
Y.~Shen, X.~He, J.~Gao, L.~Deng, and G.~Mesnil.
\newblock {A Latent Semantic Model with Convolutional-Pooling Structure for
  Information Retrieval}.
\newblock In {\em ACM conference on Conference on Information and Knowledge
  Management}, pages 101--110, New York, New York, USA, 2014. ACM Press.

\bibitem{Svore2009LTW}
K.~M. Svore and C.~J.~C. Burges.
\newblock {\em {A machine learning approach for improved BM25 retrieval}}.
\newblock ACM, New York, New York, USA, November 2009.

\bibitem{Taylor2006Optimisation}
M.~J. Taylor, H.~Zaragoza, N.~Craswell, S.~E. Robertson, and C.~Burges.
\newblock {Optimisation methods for ranking functions with multiple
  parameters.}
\newblock In {\em ACM conference on Conference on Information and Knowledge
  Management}, pages 585--593, New York, New York, USA, 2006. ACM Press.

\bibitem{Zhai2008LMOverview}
C.~X. Zhai.
\newblock {Statistical Language Models for Information Retrieval: A Critical
  Review.}
\newblock {\em Foundations and Trends in Information Retrieval}, 2(3):137--213,
  2008.

\bibitem{LeZhao2010TermNecessity}
L.~Zhao and J.~Callan.
\newblock {Term necessity prediction}.
\newblock In {\em ACM conference on Conference on Information and Knowledge
  Management}, pages 259--268, New York, New York, USA, 2010. ACM Press.

\bibitem{Zheng2015LearningToReweight}
G.~Zheng and J.~Callan.
\newblock {Learning to Reweight Terms with Distributed Representations}.
\newblock In {\em The International ACM SIGIR Conference}, pages 575--584, New
  York, New York, USA, 2015. ACM Press.

\end{thebibliography}

\end{document}